\documentclass[aip,rsi,reprint]{revtex4-2}
\usepackage[colorlinks=true,linkcolor=blue,citecolor=blue,urlcolor=blue]{hyperref}
\usepackage{graphicx}
\usepackage{amsmath}
\usepackage{siunitx}
\usepackage{xpatch} 
\usepackage{pgf}
\usepackage{tikz}
\usepackage{float}
\usepackage{soul}
\hypersetup{
    colorlinks=true,
    linkcolor=blue,
    citecolor=blue,
    urlcolor=blue
}
\makeatletter 
\renewcommand{\@biblabel}[1]{\textcolor{blue}{[#1]}} 
\makeatother
\begin{document}
\title{A Dual-Sideband Attosecond Interferometry Setup}

\date{\today}
\author{M. Jahanzeb}
\email{jahanzeb.muhammad@physik.uni-freiburg.de}
\affiliation{Physikalisches Institut, Albert-Ludwigs-Universität Freiburg, Hermann-Herder-Str. 3, 79104 Freiburg, Germany}
\author{M. Schmoll}
\email{marvin.schmoll@physik.uni-freiburg.de}
\affiliation{Physikalisches Institut, Albert-Ludwigs-Universität Freiburg, Hermann-Herder-Str. 3, 79104 Freiburg, Germany}
\author{P. Weizel}
\affiliation{Physikalisches Institut, Albert-Ludwigs-Universität Freiburg, Hermann-Herder-Str. 3, 79104 Freiburg, Germany}
\author{S. Majoni}
\affiliation{Physikalisches Institut, Albert-Ludwigs-Universität Freiburg, Hermann-Herder-Str. 3, 79104 Freiburg, Germany}
\author{R. N. Shah}
\affiliation{Physikalisches Institut, Albert-Ludwigs-Universität Freiburg, Hermann-Herder-Str. 3, 79104 Freiburg, Germany}
\author{M. Niebuhr}
\affiliation{Physikalisches Institut, Albert-Ludwigs-Universität Freiburg, Hermann-Herder-Str. 3, 79104 Freiburg, Germany}
\author{C. Manzoni}
\affiliation{Institute for Photonics and Nanotechnology - CNR, Piazza Leonardo da Vinci 32, 20133 Milano, Italy}
\author{G. Sansone}
\email{giuseppe.sansone@physik.uni-freiburg.de}
\affiliation{Physikalisches Institut, Albert-Ludwigs-Universität Freiburg, Hermann-Herder-Str. 3, 79104 Freiburg, Germany}

\begin{abstract}
We present the development and implementation of an experimental setup designed to investigate attosecond photoionization delays using a dual-sideband RABBITT (Reconstruction of Attosecond Beating By Interference of Two-Photon Transitions) technique. The setup utilizes an attosecond extreme ultraviolet source from high-harmonic generation driven by a carrier-envelope-phase-stabilized Ti:sapphire laser centered at 800\,nm. The extreme ultraviolet radiation is synchronized with a 1200\,nm infrared probe pulse generated via a non-collinear optical parametric amplifier. Active delay stabilization by means of a spectrally resolved interferometer signal achieves 45\,as root-mean-square timing precision and enables the observation of sideband oscillations. Taking advantage of the dependence of the sideband signal on the carrier-envelope phase of the driving field, we report sideband-yield oscillations as a function of this parameter.

\end{abstract}

\maketitle

\section{Introduction}

The precise measurement of attosecond delays in photoionization is critical to understanding ultrafast electron structure and dynamics in atoms, molecules, and surfaces~\cite{calegariAdvancesAttosecondScience2016,kheifetsWignerTimeDelay2023,inzaniAttosecondElectronDynamics2025}. When an atom is ionized by absorbing a photon, the wave function of the resulting photoelectron experiences a phase shift compared to the propagation of a hypothetical free electron, encoding information on the potential landscape experienced by the outgoing photoelectron wave packet~\cite{schultzeDelayPhotoemission2010, klunderProbingSinglePhotonIonization2011, maquetPhaseAndTimeDelays2014}. This phase corresponds to a time delay in photoionization, usually indicated as Eisenbud-Wigner-Smith (EWS) delay ($\tau_{\mathrm{EWS}}$)~\cite{Eisenbud1948,wignerLowerLimitEnergy1955,Smith1960}. 
Attosecond interferometry based on the measurement of photoelectron spectra in the ionization process driven by a comb of extreme ultraviolet (XUV) odd harmonics of a near-infrared (NIR) driving field and a synchronized replica of the same field, gives access to the total photoionization delay $\tau_{\mathrm{ph}}$ imposed by the interaction with the combined fields.
Under suitable approximations, this delay can be decomposed in the sum of the EWS delay and a continuum-continuum delay ($\tau_{\mathrm{cc}}$), introduced by the action of the NIR field, which determines transitions between continuum states of the photoelectron wave packet~\cite{dahlstromTheoryAttosecondDelays2012, pazourekAttosecondChronoscopyPhotoemission2015}:
\begin{equation}
    \tau_{\mathrm{ph}}=\tau_{\mathrm{EWS}}+\tau_{\mathrm{cc}}.
\end{equation}
The latter term can be interpreted as a  measurement-induced additional delay. Experimentally, attosecond interferometry based on the Reconstruction of Attosecond Beating by Interference of Two-Photon Transitions (RABBITT) technique~\cite{paulObservationTrainAttosecond2001} delivers information on the total photoionization time delay, from which the EWS delay can be derived by using a universal expression of the continuum-continuum delay derived within second-order perturbation theory~\cite{JPB-Dahlstrom-2012}. This approach necessitates a prior, quantitative characterization of the attosecond chirp associated with the XUV harmonic radiation~\cite{paulObservationTrainAttosecond2001}.

In its typical implementation, a single sideband peak is present between the main photoelectron lines generated by the absorption of XUV photons. However, recently novel configurations have been demonstrated, combining fields with different central wavelengths for the high-harmonic generation (HHG) process and for the generation of the sideband photoelectrons. 
Attosecond electron interferometry combining an attosecond pulse train driven by a NIR field with frequency $\omega_{\mathrm{NIR}}$ and radiation obtained by second harmonic of a replica of the NIR pulse has been demonstrated and used to mitigate photoelectron spectral congestion in attosecond time-delay measurements in molecules~\cite{loriotHighHarmonicGeneration2o2020}. 
Furthermore, the generation of attosecond pulse trains using a $2\omega_{\mathrm{NIR}}$ field combined with a synchronized $\omega_{\mathrm{NIR}}$ field has provided valuable information about the continuum–continuum phase in photoionization~\cite{harthExtractingPhaseInformation2019, bhartiDecompositionTransitionPhase2021, bhartiMultisidebandInterferenceStructures2024}. Finally, at the seeded free-electron laser (FEL) FERMI \cite{allariaHighlyCoherentStable2012, callegariAtomicMolecularOptical2021}, a configuration in which two sidebands between consecutive harmonics are generated has been demonstrated by using consecutive (odd and even) harmonics of a seed laser obtained by frequency triplication of a NIR field ($3\omega_{\mathrm{NIR}}$) and a replica of the latter ($\omega_{\mathrm{NIR}}$)~\cite{marojuAttosecondPulseShaping2020,marojuComplexAttosecondWaveform2021,marojuAttosecondCoherentControl2023,marojuAttosecondTemporalStructure2025}.

Attosecond interferometry experiments require highly stable interferometers that enable precise control of the delay between the XUV and NIR (or visible) pulses. The temporal delay is typically stabilized using an auxiliary laser~\cite{chiniDelayControlAttosecond2009, sabbarCombiningAttosecondXUV2014, huppertAttosecondBeamlineActively2015}, while alternatively, monolithic approaches based on collinear propagation of both pulses have been also demonstrated~\cite{ahmadiCollinearSetupDelay2020, ertelUltrastableHighrepetitionrateAttosecond2023}. Using active feedback, interferometric stabilization down to 450~as root-mean-square (RMS), including an optical parametric amplifier operating in the NIR arm, has been achieved~\cite{OE-Schlaepfer-2019}.  

In this work, we present the development of an experimental setup designed to extend attosecond interferometry to a dual-sideband configuration. In particular, the experimental setup requires a complex, active stabilization system between the pulses driving the HHG process and those responsible for the generation of the photoelectron sidebands. Furthermore, stabilization of the CEP of the driving fields, as well as spatially resolved detection of the photoelectron distribution, is required to observe oscillations of the sideband signal.
\section{Measurement Principle: Dual-Sideband RABBITT}
\begin{figure}[!htbp] 
    \centering
    \includegraphics[width=1\linewidth]{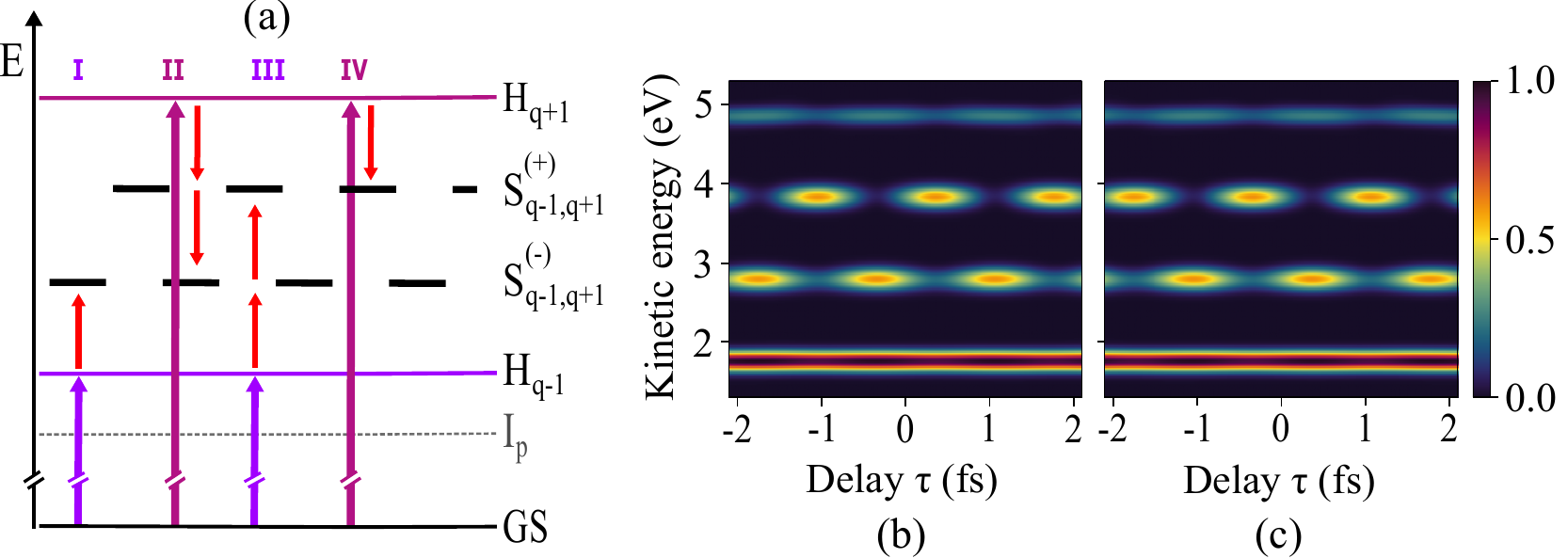}
    \caption{(a) Schematic illustration of the paths leading to the formation of the sidebands $S^{(\pm)}_{q-1,q+1}$. Adjacent odd-order, XUV harmonics (light and dark purple arrows) are coupled to the same final continuum energy by absorption or emission of one and two NIR photons (red arrows), giving rise to indistinguishable quantum pathways (I-IV). 
    (b-c) Numerical simulation of the sideband intensity oscillations as a function of the delay $\tau$ between the attosecond pulse train and the NIR field for photoelectrons emitted in opposite direction along the laser polarization direction ($\theta = 0^{\circ}$ (b), and $\theta = 180^{\circ}$ (c)).  $H_{q-1}$ and $H_{q+1}$ indicate the energy of the main photoelectron lines. GS and $I_p$ indicate the energy position of the ground state and the ionization potential of the target atom, respectively. Parameters used in the simulations: $q=18$, $I_{\mathrm{NIR}}=6\times10^{11}\,\mathrm{W/cm^2}$, target atom: Helium.}
    \label{fig:SB_Simulation}
\end{figure}

In our approach, an 800\,nm driving laser generates attosecond XUV pulses through HHG in noble gas atoms. The XUV pulses overlap with a synchronized NIR pulse with a central wavelength of 1200\,nm. This pulse is generated by a two-stage non-collinear optical parametric amplifier (NOPA) seeded by the white-light generation (WLG) created by a small fraction of the 800\,nm radiation and pumped by a fraction of the same field~\cite{cerulloUltrafastOpticalParametric2003, manzoniDesignCriteriaUltrafast2016}. The intense NIR pulse centered at 1200\,nm ($I_{\mathrm{NIR}}\approx 10^{12}$~W/cm$^2$) interacts with the photoelectrons emitted by single photon absorption from the XUV comb, leading to the emission or absorption of up to two NIR photons, which results in the formation of two sidebands between the main photoelectron peaks caused by the XUV-only absorption. Due to the ratio between the wavelengths of the NIR fields, the two sidebands (indicated as $S^{(\pm)}_{q-1,q+1}$ in Fig.~\ref{fig:SB_Simulation}a, with $q$ an even integer) generated between the main photoelectron peaks by the absorption of one photon of the harmonic $H_{q-1}$ or $H_{q+1}$ are populated by two different pathways: paths I and II for $S^{(-)}_{q-1,q+1}$ , and  paths III and IV for $S^{(+)}_{q-1,q+1}$.

These two transition routes lead to the same final continuum energy, but are characterized by the exchange of a different number of photons with the total field. 
The sideband signal emitted along the (positive) polarization direction ($\theta=0^{\circ}$) of the XUV and NIR fields is expressed by the relation~\cite{paulObservationTrainAttosecond2001}:

\begin{equation}\label{Eq3}
    \begin{split}
        S^{(\pm)}_{q-1,q+1}(\tau)&=a_{q-1,q+1}\pm b_{q-1,q+1}\times \\&\ \times\cos\left[3\omega_{\mathrm{NIR}}(\tau+\tau_{\mathrm{ph}})+\left(\varphi^{(0)}_{q+1}-\varphi^{(0)}_{q-1}\right)+\varphi_{\mathrm{CEP}}\right],  
    \end{split}
\end{equation}
where the constants $a_{q-1,q+1}$ and $b_{q-1,q+1}$ depend on the photoelectron energy and NIR intensity~\cite{marojuAnalysisTwocolorPhotoelectron2021}. 
The term $\varphi_{\mathrm{CEP}}$ indicates the carrier-envelope phase of the driving field centered at 800\,nm; the phases $\varphi_{q\pm1}$ of the two harmonics $H_{q-1}$ or $H_{q+1}$ are given by the expression:
$\varphi_{q\pm1}=\varphi^{(0)}_{q\pm1}+(q\pm1)\varphi_{\mathrm{CEP}}$~\cite{PRL-Sansone-2005,sansoneControlLongElectron2006}.
It is important to observe that the sideband signal emitted in the opposite direction along the laser polarization ($\theta=180^{\circ}$) is obtained by interchanging the sign of the oscillating terms in Eq.~\ref{Eq3}. As a result the oscillations of the photoelectron yield for the same sideband emitted in opposite directions along the polarization direction oscillate with a $\pi$ phase offset (see Fig.~\ref{fig:SB_Simulation}b and c) and the overall yield does not depend on the XUV-NIR delay. 
The phase opposition of the oscillations for the two opposite detection directions $\theta=0^{\circ}$ and $\theta=180^{\circ}$ is clearly demonstrated by the simulations based on the strong-field-approximation model~\cite{kitzlerQuantumTheoryAttosecond2002, itataniAttosecondStreakCamera2002}, shown in Fig.~\ref{fig:SB_Simulation}b and Fig.~\ref{fig:SB_Simulation}c, respectively. 
This property is related to the observation that the two paths contributing to the same sidebands are characterized by the exchange of a different number of photons, leading to final states with different parity. As a result, for each sideband, when integrating over the total solid angle, the total photoelectron yield does not change as a function of the relative delay between the XUV and NIR pulses. However, angular-resolved detection of the photoelectron spectra allows one to observe a beating of the signal at $3\omega_{\mathrm{NIR}}$, when limiting the signal integration to only a semi-volume (or part of it) along the laser polarization direction. 

Equation~\ref{Eq3} indicates that the sideband oscillations depend on the CEP of the driving field. This property is peculiar of our scheme and indicates that observation of the sideband oscillations does not only require stabilization of the relative delay between the XUV and NIR pulses, but also stabilization of the CEP of the two NIR fields. For the derivation of Eq.~\ref{Eq3}, we assume ideal conditions in which the CEP drift of the NIR pulse generated by the NOPA is identical to that of the 800\,nm driving field used for high-harmonic generation~\cite{PRL-Baltuska-2002}. 

\section{Experimental Setup}

The experiments were carried out using a CEP stabilized Ti:sapphire laser system delivering 30~fs pulses with an energy of 3~mJ at a repetition rate of 1~kHz. The residual CEP jitter is below 200~mrad (RMS)~\cite{luckingApproachingLimitsCarrier2014,luckingLongtermCarrierenvelopephasestableFewcycle2012,kokeDirectFrequencyComb2010}.

\subsection{XUV spectrum}
The laser pulses were split into two arms using a 50:50 beam splitter.
The pulses in the first arm were used to drive the HHG process in argon, as depicted in Fig.~\ref{fig:Full_setup}. The estimated intensity of the driving pulse in the gas jet was $\sim 2\times10^{14}\,\mathrm{W/cm^2}$. The XUV harmonics were then refocused with unitary magnification by a toroidal mirror into the interaction region of a velocity map imaging spectrometer (VMI).

\subsection{Generation and characterization of NOPA pulses}
\begin{figure*}[t]
    \centering
    \includegraphics[width=0.7\linewidth]{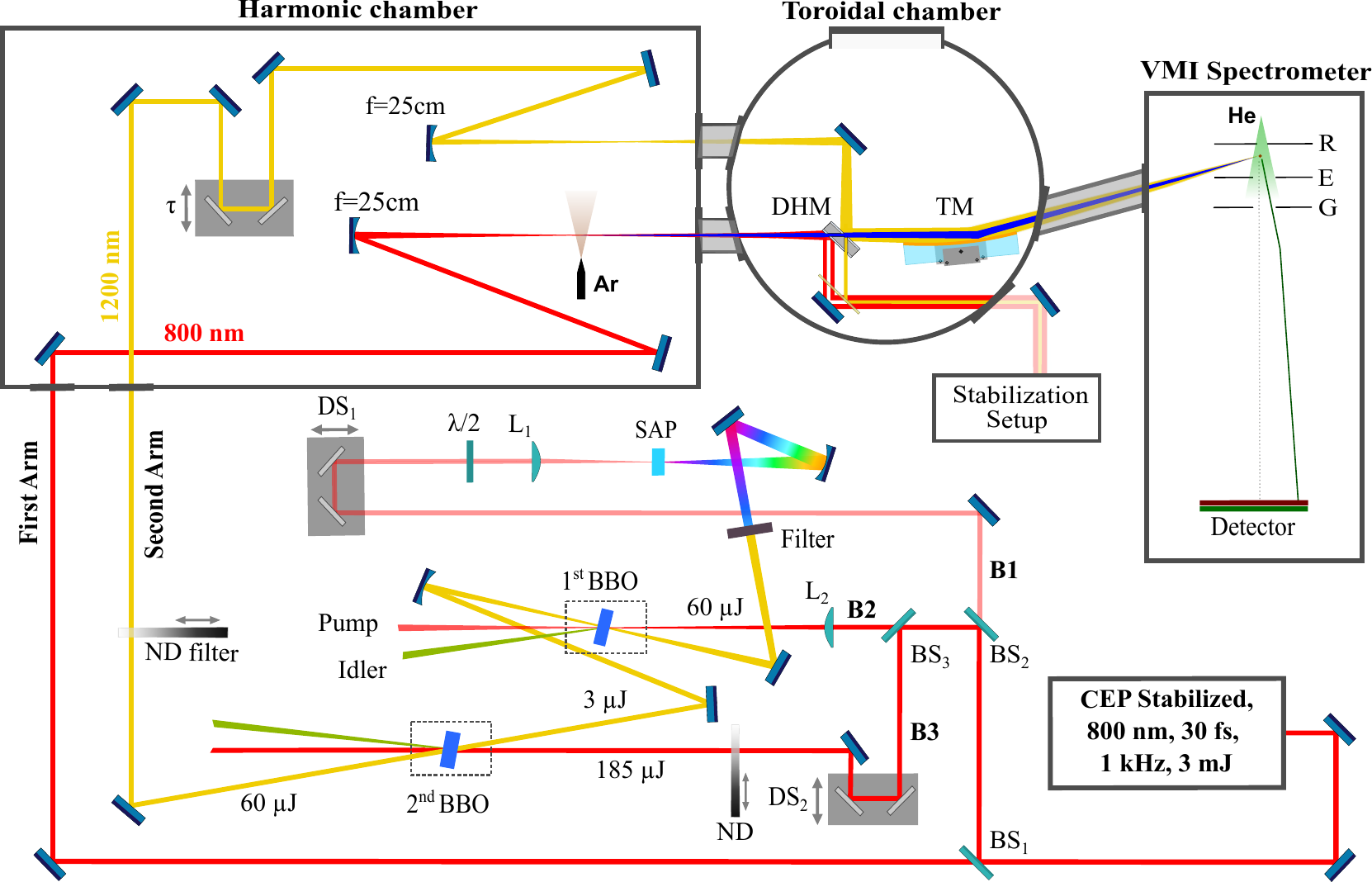}
    \caption{Schematic of the complete experimental setup. The laser output is split by a 50:50 beam splitter ($\mathrm{BS}_1$) into two arms. The first arm is directed to the HHG chamber, where the 800\,nm driving pulse is focused into an argon gas jet to generate an attosecond XUV pulse train. The second arm drives a two-stage NOPA. Within the NOPA, the beam is further divided into three branches by using the beam splitters $\mathrm{BS}_2$ and $\mathrm{BS}_3$, respectively. The first branch (B1) generates a white-light continuum in a sapphire crystal (SAP) for seeding the NOPA stages; a long-pass filter placed after the WLG stage suppresses unwanted spectral components and is removed when operating the 800\,nm stabilization scheme. $L_1$ indicates the lens used to focus the 800\,nm radiation in the sapphire plate. The remaining two branches B2 and B3 serve as pump beams for the first and second NOPA stages, respectively. In the branch B2, the 800\,nm pump is focused by the lens $L_2$ in the BBO crystal. A variable neutral density (ND) filter is placed immediately after the NOPA to tune the NIR intensity. The XUV and NIR pulses are recombined using a double-holey mirror (DHM) with reflective coating on both surfaces and focused by a toroidal mirror (TM) in the interaction region of a VMI spectrometer, consisting of the three electrodes repeller (R), extractor (E) and ground (G) that guide the photoelectrons towards the detector. A small fraction of the co-propagating beams is extracted through the DHM and routed to the delay stabilization setup for active interferometric phase locking. The delay stages $DS_1$, $DS_2$ were used to optimize the temporal overalp required for efficient amplification in the first and second NOPA stage. The piezo stage $\tau$ was used to optimize, stabilize and control the delay $\tau$ between the XUV and NIR pulses in the VMI region.
}
    \label{fig:Full_setup}
\end{figure*}

The second part of the beam was used for the generation of ultrashort laser pulses centered around 1200\,nm, employing a two-stage NOPA, as illustrated in Fig.~\ref{fig:Full_setup}. The seed for the parametric amplifiers was produced via WLG by focusing the fundamental 800\,nm driving laser into a 3-mm-thick sapphire plate. Through third-order nonlinear interactions, the white light spans from the visible to the mid-infrared spectral region. The spectral portion of the continuum around 1200\,nm is selectively amplified by non-collinear parametric amplification in beta-barium borate (BBO) crystals.  
To reach the pulse intensities required for the exchange of two NIR photons in the photoionization experiments, two amplification stages are implemented. In the first stage, a 0.7-mm-thick type-II BBO crystal produces pulses with an energy of about 3~$\mathrm{\mu}$J. The second stage, employing a thicker (1.4~mm) BBO crystal, increases the output by approximately a factor of 20, delivering NIR pulses with energies of ~60~$\mathrm{\mu}$J.

\begin{figure}[htbp]
    \centering
    \includegraphics[width=1\linewidth]{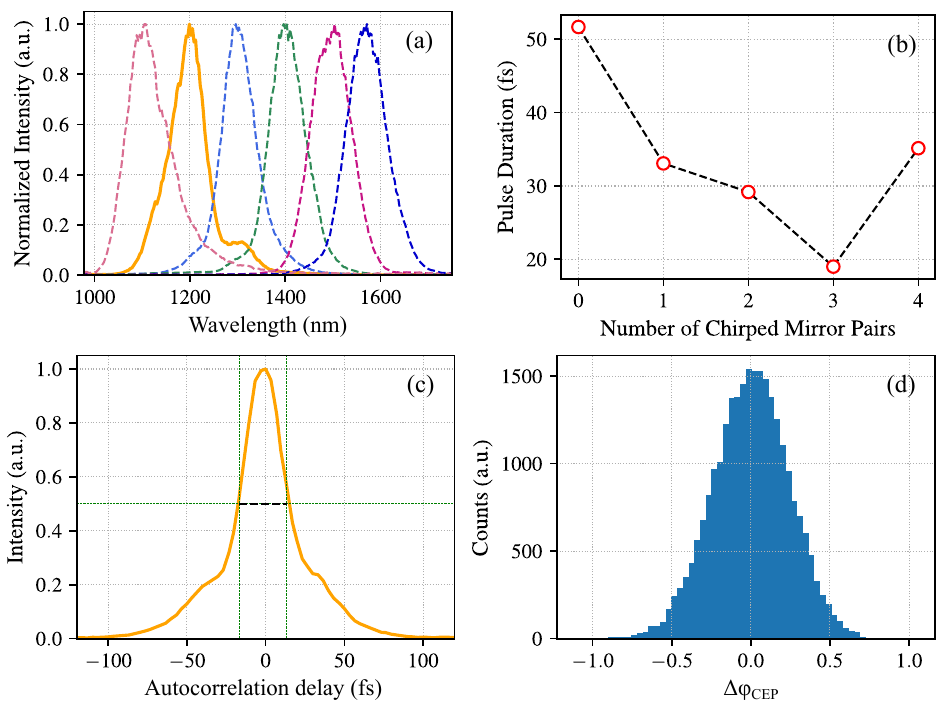}
     \centering
     \caption{(a) Normalized NOPA spectrum at the output of the second amplification stage demonstrating the central wavelength tunability from approximately 1100~nm to 1700~nm. (b) Pulse duration as a function of the number of chirped mirror pairs used for dispersion compensation. (c) Second-order autocorrelation trace recorded with three pairs of chirped mirrors demonstrating temporal compression down to approximately 19~fs (FWHM), close to the Fourier-transform limit (18~fs). (d) CEP stability of the NOPA obtained from an $f$--$2f$ spectral interferometer, yielding an RMS CEP phase jitter of 244~mrad. }
    \label{fig:NOPA_Tunability}
\end{figure}

The non-collinear configuration supports a broad gain bandwidth, enabling the generation of widely tunable pulses with broad spectral bandwidth. Indeed, by rotating the BBO crystal, one can adjust the phase-matching condition and thus amplify different spectral regions of the white-light continuum. This allows the generation of infrared pulses over a broad wavelength range, from approximately 1100~nm to 1700~nm, as presented in Fig.~\ref{fig:NOPA_Tunability}a. Such flexibility makes the NOPA a versatile source for experiments requiring wavelength-dependent studies or specific photon energy, depending on the system under investigation.  

Based on the measured spectral bandwidth of the NOPA output, the Fourier-transform limit (FTL) of the pulses was estimated to be approximately 18~fs. To approach this limit, dispersion compensation was implemented using chirped mirrors (PC94, Ultrafast Innovations). Broadband chirped mirror pairs were selected to compensate for the accumulated group-delay dispersion (GDD), while minimizing higher-order phase distortions~\cite{steinmeyerBrewsterAngledChirpedMirrors2009}.
The amplified pulses from the NOPA were characterized using a second-order autocorrelation setup to retrieve information about their temporal intensity profile~\cite{trebinoUsingPhaseRetrieval1997}. The dependence of the measured pulse duration for different pairs of chirped mirrors is shown in Fig.~\ref{fig:NOPA_Tunability}b. 
After three pairs of chirped mirrors, the pulse duration was reduced to 19~fs, which is close to the calculated FTL, as presented in Fig.~\ref{fig:NOPA_Tunability}c. 

An important aspect of our experimental apparatus is that the WLG inherits the CEP drift of the driving laser, ensuring that the seed of the NOPA is CEP-stable. Since optical parametric amplification preserves the CEP drift of the seed laser~\cite{PRL-Baltuska-2002}, the amplified probe pulses are expected to maintain the CEP-stability. However, intensity fluctuations in nonlinear crystals may lead to nonlinear intensity–phase coupling, introducing additional phase noise~\cite{a.baltuskaPhasecontrolledAmplificationFewcycle2003}. 
We therefore characterized the CEP stability of the final NOPA output using an $f$--$2f$ interferometer based on supercontinuum generation in sapphire and second-harmonic generation in BBO. The fringe phase is extracted from each recorded spectrum using an FFT-based analysis and after removal of slow interferometer drifts, we obtain an RMS CEP phase jitter of $\sigma_{\varphi_{\mathrm{CEP}}}=244$\,mrad for the 1200\,nm pulses when the Ti:sapphire source was CEP stabilized (Fig.~\ref{fig:NOPA_Tunability}d), confirming that the NOPA delivers CEP-stable probe pulses suitable for the phase-sensitive measurements presented below.

\subsection{Delay Stabilization}

The observation of sideband oscillations in attosecond interferometry requires attosecond-level stabilization of the relative delay between the XUV and NIR fields. This imposes stringent constraints on the temporal stability of the interferometer. The optical beam path in our experimental setup has a total length of approximately 8~m and exhibits short-term delay fluctuations $\Delta\tau$ of 520~as, corresponding to phase jitter $\Delta\phi=3\omega_{\mathrm{NIR}}\Delta\tau$ ($\omega_{\mathrm{NIR}}=800$~nm) fluctuations of about 1.22~rad (RMS). 

\begin{figure}[htbp]
    \centering
    \includegraphics[width= 0.8\linewidth]{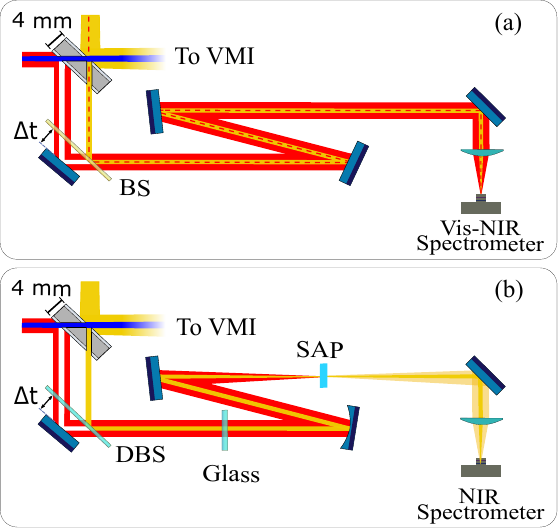}
    \caption{
    Recombination of the two interferometer arms using a double-coated, double-holey mirror.
    Panel (a) shows the 800\,nm stabilization scheme, while panel (b)  shows the 1200\,nm stabilization scheme, which requires a sapphire (SAP) crystal for WLG. The red and dark yellow lines indicate the radiation at 800\,nm and 1200\,nm, respectively. BS: beam splitter; DBS: dichroic beam splitter.
    }
    \label{fig:Beam_combining_setup}
\end{figure}

The XUV pump and NIR probe arms are recombined using a double-coated, double-holey mirror, as depicted in Fig.~\ref{fig:Beam_combining_setup}. 
The front surface of the mirror presents a single central hole, which is the starting point of two channels in the mirror substrate (diameter of the channels is 2\,mm) oriented at 45$^{\circ}$ with respect to the surface normal and at 90$^{\circ}$ with respect to each other. The exits of the two channels define two off-centered holes on the back surface of the mirror.
The front surface of the mirror reflects an annular section of the 1200\,nm radiation and directs it towards the interaction region of the VMI, where it is spatially and temporally overlapped  with the XUV radiation propagating through one of the channels in the holey-mirror substrate.

The central part of the 1200\,nm radiation is transmitted through the second channel of the holey mirror.
The back surface of the mirror reflects an annular section of the 800\,nm co-propagating with the XUV radiation. These two beams are used to stabilize the delay in the interferometer by spectral interferometry. However, due to the finite thickness of the drilled mirror $d=4\,\mathrm{mm}$ and the 45$^\circ$ angle of incidence, the beams are laterally displaced and present a path-length mismatch of $\frac{\sqrt{2}d}{c}$.

To address this, we designed a compact beam-combining stage. It consists of a 1-inch-diameter broadband mirror and a rectangular beam splitter/dichroic mirror (25\,mm~$\times$~36\,mm) mounted on a precision micro-translation stage.
As shown in Fig.~\ref{fig:Beam_combining_setup} this setup allows the 800\,nm beam to travel additional distance, leading to a delay $\Delta t$, which can be adjusted via the micro-positioning stage.
Finally, the stabilization scheme is designed to be versatile, as the interferometer can be actively locked using pulses centered either at 800\,nm or at 1200\,nm, enabling flexible operation under different experimental conditions.

\subsubsection{Stabilization Scheme at 800\,nm}
In this configuration, the delay stabilization is performed using the fundamental Ti:sapphire wavelength at 800\,nm in both interferometer arms. Since the second arm contains a residual fraction of the fundamental that co-propagates with the NOPA output, no additional nonlinear conversion is required to generate the stabilization signal. This residual 800\,nm component originates from the fact that the NOPA is seeded by the WLG continuum. Although the 800\,nm component could be removed by filtering, it is intentionally preserved to enable the stabilization of the interferometer using radiation at this wavelength.

For this configuration, a plate beam splitter with 90\% reflectivity and 10\% transmission is used after the double-holey mirror to preferentially transmit the weak 800\,nm beam from the first arm of the interferometer, while strongly reflecting the 800\,nm beam that is co-propagating in the NOPA output, as shown in Fig.~\ref{fig:Beam_combining_setup}a. Both collinear beams are routed out of the vacuum chamber and directed to the stabilization setup. No white-light generation is required in this configuration. The spectral interference fringes are recorded using a visible-NIR spectrometer, which provides the signal for active delay stabilization.

The delay control and stabilization are characterized by performing two complementary types of measurements: (i) a controlled delay scan for the acquisition of RABBITT traces, and (ii) long-term fixed-delay locking to assess stability over extended durations.
During the first type of measurement, the feedback loop was commanded to follow a linear phase ramp from $0$ to $6\,$rad over a period of five minutes. During this acquisition, the measured phase has a residual fluctuation of $\sigma_{\phi} = 0.11\,$rad, corresponding to a timing jitter of $\sigma_{\tau} = 45.16\,$as (RMS). This level of performance outperforms stabilization results reported in earlier interferometric control schemes using radiation with different central wavelengths in the pump and probe arm of attosecond interferometers~\cite{OE-Schlaepfer-2019}.

\begin{figure}[h!]
    \centering
    \includegraphics[width=1\linewidth]{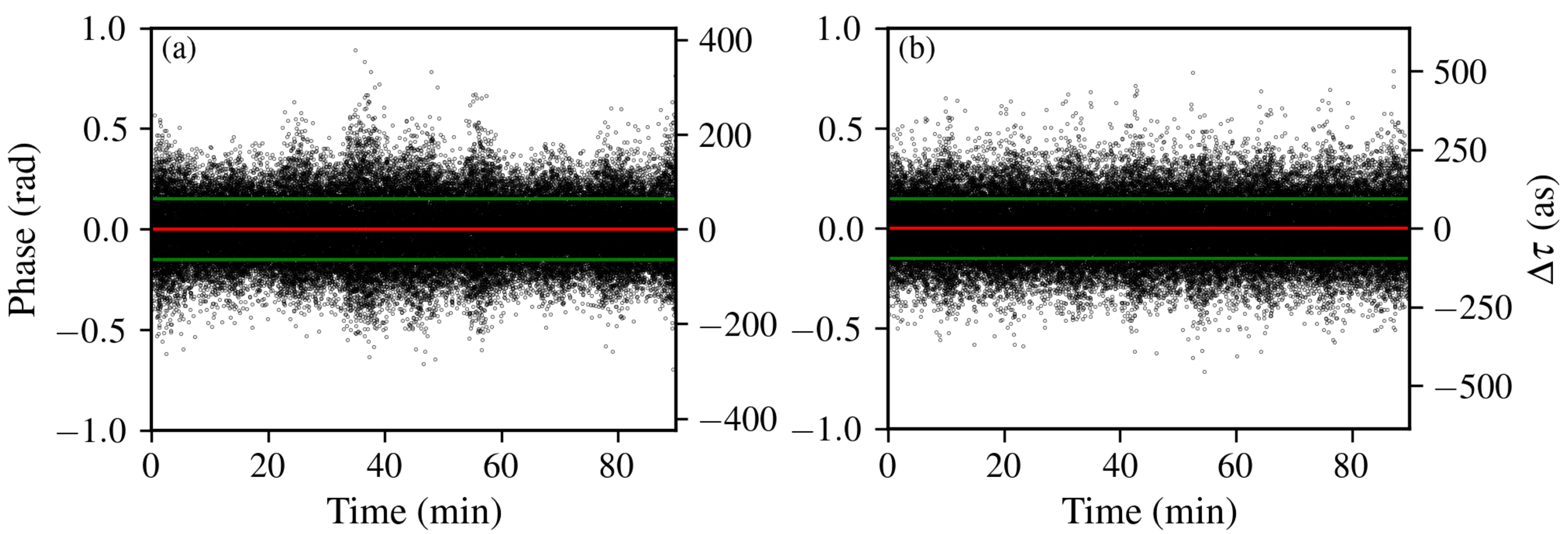}
    \caption{Long-term delay stabilization performance for the 800\,nm and 1200\,nm schemes. The interferometer was locked to a fixed phase setpoint (red line) and monitored over 90\,min. The phase stability corresponded to an RMS timing jitter of 64.18 as for the 800\,nm scheme (panel a) and 94.76 as for the 1200\,nm scheme (panel b). The green lines indicate the $\pm\sigma_{\tau}$ stability bounds.
}
    \label{fig:800long}
\end{figure}

Long-term stability is shown in Fig.~\ref{fig:800long}a, where the interferometer was locked to a fixed phase setpoint for $90$\,minutes. The delay stability was maintained with $\sigma_{\phi} = 0.15$\,rad ($\sigma_{\tau} = 64.18$\,as).

These results demonstrate that the delay stabilization approach achieves attosecond timing precision over both dynamic (phase-scanned) and static (fixed-phase) operation modes.

\subsubsection{Stabilization Scheme at 1200\,nm}
In the 1200\,nm configuration, the active stabilization is performed at the central NOPA wavelength. In this scheme, the same optical setup from the previous configuration is used.
To access the 1200\,nm component for stabilization, a dichroic short-pass beam splitter (DBS) is used in the recombination stage, instead of a beam splitter. This optic transmits the 800\,nm radiation, while reflecting the 1200\,nm NOPA beam. After collinear combination using the custom mirror–DBS stage, the beam is routed out of the vacuum chamber and is focused into a 3\,mm sapphire plate for WLG, as shown in Fig.~\ref{fig:Beam_combining_setup}b. The spectral continuum contains sufficient energy density around 1200\,nm to interfere with the NOPA output. Because the white-light is highly divergent, it is re-focused, and interference fringes are recorded using a NIR spectrometer. To easily optimize the relative delay, a 3\,mm glass plate is inserted before the WLG. Its dispersion introduces an effective wavelength-dependent delay, affecting the 800\,nm and 1200\,nm components differently and improving the spectral interference fringes used for stabilization.

This delay stabilization scheme was also tested in the two different operation modes described in the previous section.
During the short scan the phase is ramped from 0\,rad to 6\,rad with a residual phase fluctuation of $\sigma_{\phi} = 0.113$\,rad, corresponding to $\sigma_{\tau} = 72.22$\,as RMS over a period of five minutes. Long-term stability measurements are shown in Fig.~\ref{fig:800long}(b), where the interferometer was locked to a constant phase setpoint for 90\,minutes. The residual phase jitter remained at $\sigma_{\phi} = 0.149$\,rad, equivalent to $\sigma_{\tau} = 94.76$\,as, without any observed loss of tracking. The performance is sufficient for extended attosecond pump–probe acquisitions, improving the stability achieved in optical delay stabilization systems of similar setups~\cite{OE-Schlaepfer-2019,chiniDelayControlAttosecond2009}. 

Both stabilization schemes are fully interchangeable, and the mechanical design of the beam-combining module allows switching between 800\,nm and 1200\,nm operation without realignment of the main beam path. This dual-wavelength stabilization capability adds flexibility to the setup and enables optimization of interferometric stability under different experimental conditions (for example, different harmonic generation conditions and driving wavelengths). The results obtained with the two stabilization schemes are summarized in Table~\ref{tab:stabilization_comparison}.
We can observe that both stabilization schemes provide attosecond-level delay control
for both dynamic and static operation modes. 

\begin{table}[h]
\centering
\caption{Comparison of the two active delay stabilization schemes. 
The table summarizes the operating wavelength, stabilization method, and achieved phase and delay stability for the dynamic mode (DM, phase scan) and static mode (SM, long-term lock).}
\label{tab:stabilization_comparison}
\begin{tabular}{lcc}
\hline\hline
\textbf{Parameter} & \textbf{800\,nm} & \textbf{1200\,nm} \\
\hline
Phase range & $0$--$6$ rad & $0$--$6$ rad \\
Phase step size & $0.1$ rad & $0.1$ rad \\
DM phase jitter (RMS) & $0.11$ rad & $0.11$ rad \\
DM delay jitter (RMS) & $45.16$ as & $72.22$ as \\
DM stabilization time & 5 min & 5 min \\
\hline
SM phase jitter (RMS) & $0.15$ rad & $0.15$ rad \\
SM delay jitter (RMS) & $64.18$ as & $94.76$ as \\
SM stabilization time & 90 min & 90 min \\
\hline\hline
\end{tabular}
\end{table}

\subsection{Photoelectron Mapping Using VMI}
The photoelectron spectrometer used in the experiment is a VMI spectrometer~\cite{eppinkVelocityMapImaging1997} combined with a pulsed gas nozzle operating at repetition rates up to 1~kHz. The pulsed nozzle provides high target gas densities in the interaction region, while maintaining a low average gas load in vacuum. Under optimal operating conditions  with a backing pressure $\sim$20\,bar, the pulsed nozzle increased the total photoelectron signal in argon by nearly a factor of $\sim$6 compared to a continuous nozzle. The enhanced target density provided by the pulsed nozzle was key to enabling high-quality two-sideband RABBITT measurements. 
In order to account for inhomogeneity of the detector-response, we acquire high-statistics XUV+NIR calibration measurements that were used to correct the slightly asymmetric response of the upper and lower halves (with respect to the laser polarization direction) of the detector.

\section{Results and Discussion}
According to Eq.~\ref{Eq3}, the sideband intensity depends on both the relative phase between the attosecond pulse train and the NIR field, and on the CEP of the latter. As a result, sideband oscillations can be measured either by changing the relative delay $\tau$ between the two pulses characterized by a stable CEP, or by fixing the relative delay between the XUV and NIR field and changing the CEP of the driving field. We acquired measurements in both ways by implementing the 1200\,nm stabilization scheme.

\begin{figure}[h]
    \centering
    \includegraphics[width=\columnwidth]{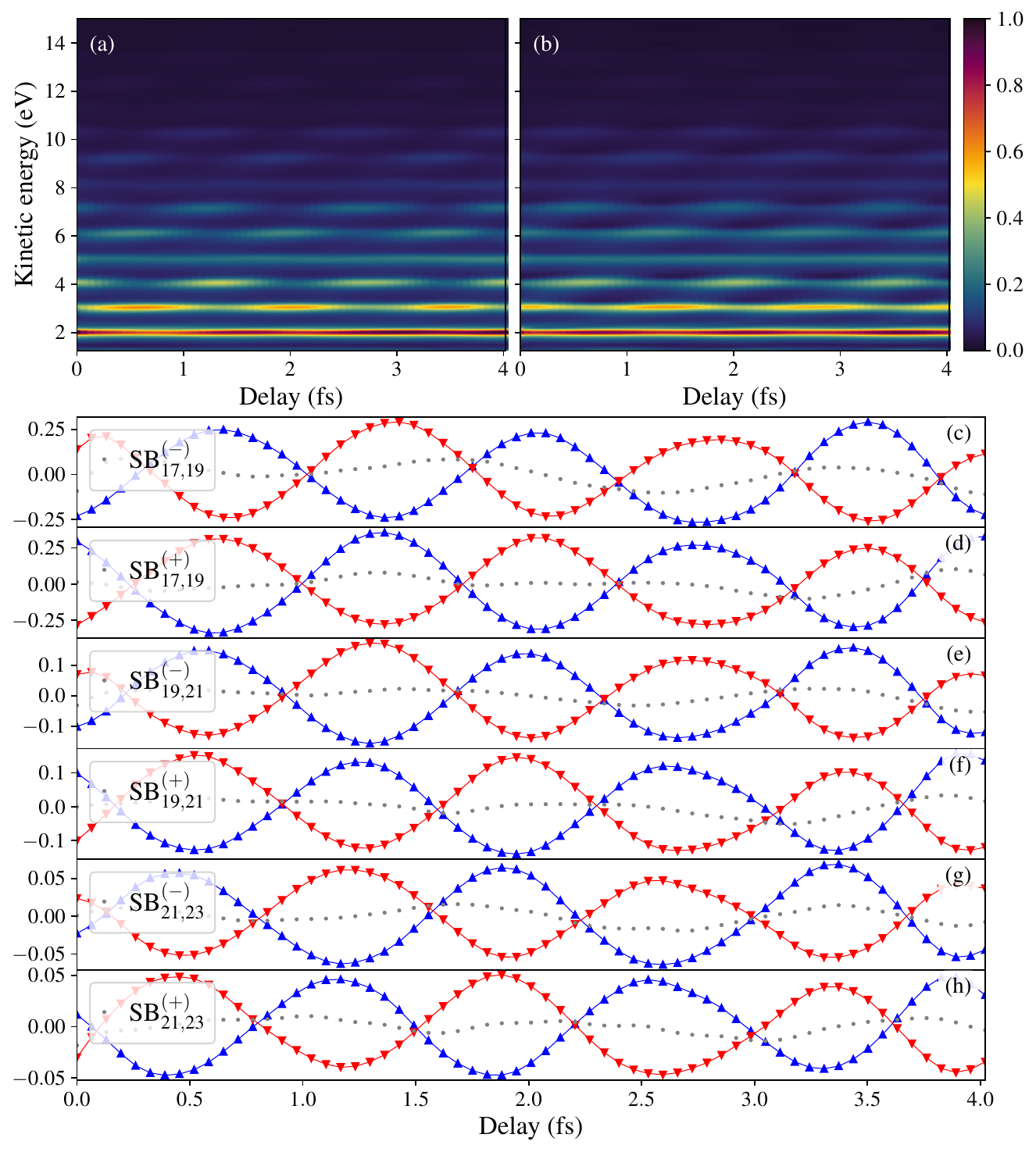}
    \caption{(a-b) RABBITT traces measured in helium by scanning the relative delay between the attosecond XUV pulse train and synchronized NIR probe pulse (6.1$\,\mu\mathrm{J}$) . The photoelectron signals are integrated over the entire upper (a) and lower (b) semi spaces with respect to the laser polarization direction. (c-h) Integrated sideband signals as a function of the relative delay $\tau$ for the upper (blue) and lower (red) semi spaces: $S^{(-)}_{17,19}$ (c), $S^{(+)}_{17,19}$ (d), $S^{(-)}_{19,21}$ (e), $S^{(+)}_{19,21}$ (f), $S^{(-)}_{21,23}$ (g) and $S^{(+)}_{21,23}$ (h). The gray curves are the sum of the sideband signals measured in the two semi spaces. All sideband signals are relative to the corresponding average over all delays.
}
    \label{fig:SB_Data}
\end{figure}

The measurement acquired by changing the relative delay $\tau$ is presented in Fig.~\ref{fig:SB_Data}. During the experiment, the interferometer was actively stabilized using spectral fringes originating from 1200\,nm--1200\,nm interference. The probe pulse energy was kept constant at 6.1\,\(\mu\)J throughout this measurement. The phase delay was swept from 0 to 6.3\,rad (corresponding to a temporal scan of $\sim$4.03\,fs) in steps of 0.1\,rad (i.e.\ $\sim$64\,as per step). Clear delay-dependent oscillations of the two-photon sidebands are visible. The CEP of the 800\,nm radiation was stabilized during the measurement with a residual RMS of 200\,mrad.
The sideband signals oscillate as a function of the relative delay with a period equal to $T=1.33\,\mathrm{fs}$, according to Eq.~\ref{Eq3}. Furthermore, the oscillations of the sideband present a $\pi$ phase offset between the upper (blue lines) and lower (red lines) semi-volumes, due to the different parity of the final wave function in the continuum reached following the two different pathways. As anticipated, the total yield of each sideband integrated over the entire volume (grey lines) does not significantly depend on the relative delay $\tau$.

\begin{figure}[h]
    \centering
    \includegraphics[width=1\linewidth]{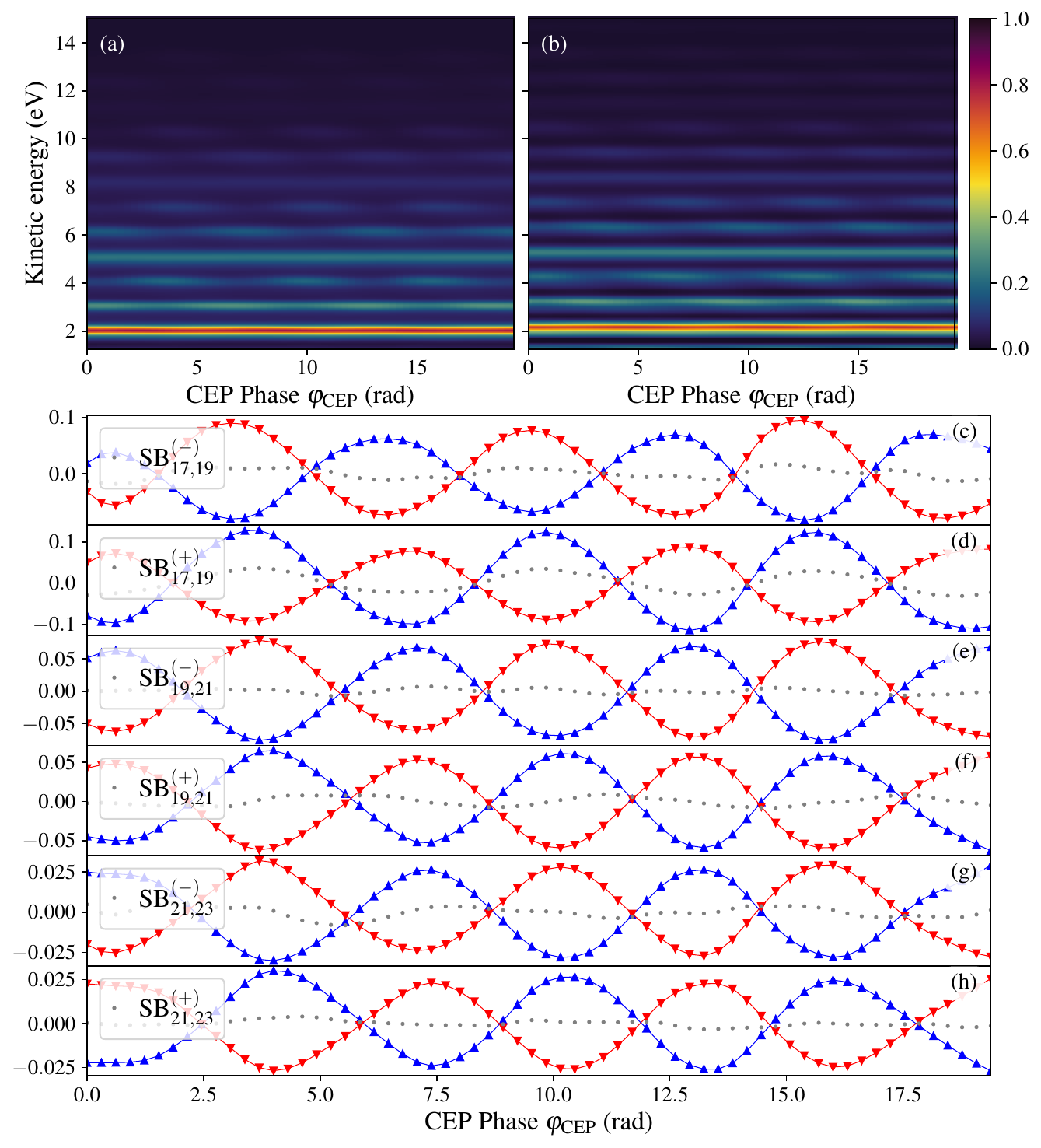}
    \caption{(a-b) RABBITT traces  measured in helium by changing the CEP of the 800\,nm driving pulse. The energy of the 1200\,nm NIR pulse is (3.7$\,\mu\mathrm{J}$). The photoelectron signals are integrated over the entire upper (a) and lower (b) semi spaces with respect to the laser polarization direction. (c-h) Integrated sideband signals as a function of the CEP phase $\varphi_{\mathrm{CEP}}$ for the upper (blue) and lower (red) semi spaces for the sidebands: $S^{(-)}_{17,19}$ (c), $S^{(+)}_{17,19}$ (d), $S^{(-)}_{19,21}$ (e), $S^{(+)}_{19,21}$ (f), $S^{(-)}_{21,23}$ (g) and $S^{(+)}_{21,23}$ (h). The gray curves are the sum of the sideband signals measured in the two semi spaces. All sideband signals are relative to the corresponding average over all CEP phases.}
    \label{fig:SB_CEP_Data}
\end{figure}

A second measurement was carried out by scanning the CEP while keeping the relative XUV–NIR delay fixed, as shown in Fig.~\ref{fig:SB_CEP_Data}. For this scan, the probe pulse energy was deliberately lowered to 3.7\,\(\mu\)J using the variable ND filter (see Fig.~\ref{fig:Full_setup}), demonstrating the capability to acquire scans at different probe intensities. The CEP variation was introduced by translating a pair of fused silica wedges (apex angle $4^{\circ}$) in steps of 40~$\mu$m, resulting in a phase change of approximately 0.3~rad per step. Over a total translation distance of 2.5~mm, a cumulative CEP shift of approximately 6$\pi$~rad was achieved. During this acquisition, the XUV–NIR delay stabilization exhibited a residual phase jitter of 110 mrad, equivalent to $\sim$70~as RMS.

This measurement reveals a clear oscillatory modulation of the sidebands as a function of CEP, validating the dual-sideband RABBITT response under controlled CEP variation.

Similar results were achieved using the stabilization based on the 800\,nm radiation. Furthermore, we have verified that without CEP stabilization, no oscillations of the sideband could be observed. This is consistent with Eq.\ref{Eq3} in which the average over the CEP leads to a vanishing oscillation of the sideband intensity measured along the positive (or negative) polarization direction of the fields.

\section{Conclusion}

 We demonstrated a compact lab-based, dual-sideband attosecond interferometry setup with (i) tunable and CEP-stable NOPA probe, (ii) interchangeable 800\,nm and 1200\,nm stabilization with down to 45\,as RMS delay jitter, and (iii) pulsed-nozzle VMI for high signal-to-noise ratio, angle-resolved detection. The system reproduces dual-sideband conditions in a lab environment and is suitable for phase-resolved, energy and angle-dependent delay measurements. Compared to FEL-based setups, this implementation is compact and scalable for broader use in attosecond science. The system will be used in the future investigation of continuum-continuum effects predicted to occur when using multisideband configurations.

\section*{DATA AVAILABILITY}
The data that support the findings of this study are available on Zenodo: \href{https://doi.org/10.5281/zenodo.19163732}{10.5281/zenodo.19163732}.

\begin{acknowledgments}

We acknowledge the financial support from the Deutsche Forschungsgemeinschaft project Research Training Group DynCAM (RTG 2717) and from the European Union's Horizon Europe research and innovation programme under the Marie Skłodowska-Curie grant agreement No 101168628 (project QU-ATTO).
We thank the group of Prof. Frank Stienkemeier (University of Freiburg) for providing us with a CRUCS valve used in the present experiments.
\end{acknowledgments}
\bibliography{aipsamp}
\end{document}